\documentstyle[amscd,amssymb,verbatim,12pt]{amsart}
\def\dspace{\baselineskip=0.3 in}
\begin{document}
\dspace
\title[Cosmic evolution ......]{COSMIC EVOLUTION WITH EARLY AND LATE
  ACCELERATION INSPIRED BY DUAL NATURE OF THE RICCI SCALAR CURVATURE }

\author[S.K.Srivastava]%
        {    }

\maketitle

\centerline{\bf S.K.Srivastava }

\centerline{ Department of Mathematics, North Eastern Hill University,}

\centerline{ NEHU Campus,Shillong - 793022 ( INDIA ) }

\centerline{e-mail:srivastava@@nehu.ac.in ; sushilsrivastava52@@gmail.com }

\vspace{1cm}

\centerline{\bf Abstract}

\smallskip

In the present paper, it is found that dark energy emerges spontaneously from
modified gravity. According to cosmological scenario, obtained here, the
universe inflates for $\sim 10^{-37} {\rm sec.}$ in the beginning and late
universe accelerates after $8.58 {\rm Gyrs}$. During the long
intermediate period, it decelerates driven by radiation and
subsequently by matter. Emerged gravitational dark energy mimics quintessence
and its density falls by 115
orders from its initial value $2.18 \times 10^{68} {\rm GeV}^4$ to current
value $2.19 \times 10^{-47} {\rm GeV}^4$. 


\newpage

\centerline{\bf 1. Introduction}

Recent astronomical observations have revolutionized the
understanding of 
cosmology. Consequences of combined analysis of Ia Supernovae (SNe Ia) observations
\cite{sp,ar}, galaxy cluster measurements \cite{na} and cosmic microwave
background (CMB) data \cite{dn} has shown that dark energy dominates the
present universe causing cosmic acceleration. This acceleration is observed at a
very small red-shift showing that it is a recent phenomenon in the late
universe. Observations of 16 type Ia Supernovae made by Hubble Space Telescope
(HST)\cite{ag}, modified earlier astronomical results and provided conclusive
evidence for deceleration prior to cosmic acceleration caused by dark energy in the recent
past.

 The acceleration is realized with negative pressure and positive energy
density that violate strong energy condition (SEC) \cite{brp}. This violation of SEC gives reverse
gravitational effect. Due to this effect, universe gets a jerk and
transition from deceleration to acceleration takes place. Of
late, phantom energy has appeared as a potential dark energy candidate in this arena, which
violates weak as well as strong energy condition \cite{rr,sk}.  

Although dark energy dominates today
, its present density is very low $\sim 2.18 \times
10^{-47} {\rm GeV}^4$. The dark energy density, in the early universe, is found to have been very high. For
example, the density at the Planck scale is found to be $1.18 \times 10^{75} {\rm GeV}^4$,
at the GUT (grand unified theory) scale it is $\sim 10^{23} {\rm GeV}^4$ and at
QCD scale it is $\sim 10^{-6} {\rm GeV}^4.$ 

The simplest dark energy candidate is the
cosmological constant $\Lambda$, but it needs to be extremely fine-tuned to satisfy the current value of the dark energy density. Alternatively, to explain decay of
the density, many dynamical
models have been suggested, where $\Lambda$ varies slowly with cosmic time ($t$) \cite{jm, vs}. In addition to models with dynamical $\Lambda (t)$, many hydrodynamic models
with or without dissipative pressure have been proposed in which barotropic fluid is
the source of dark energy \cite{vs}.  Chaplygin gas as well as generalized Chaplygin gas have also been considered as possible dark energy sources due to negative pressure of these \cite{sk, rj, ob, mc}. 

In field-theoretic models, scalar field sources of dark energy are the most natural. Apart from quintessence and
phantom \cite{brp, rr,sk}, other scalar field models involve k-essence \cite{ca}
and tachyon fields\cite{as}. Ref. \cite{ms} proposes dark energy models derived
from the action with string gravity corrections as well as dilaton and modulus
fields.  

Other than these approaches, some authors have
considered modified gravitational action by adding a function $f(R)$ ($R$
being the Ricci scalar curvature) to Einstein-Hilbert lagrangian, where $f(R)$
provides a gravitational alternative for dark energy causing late-time
acceleration of the universe. In \cite{sc,sm}, $f(R) = 1/R$, but this has the
problem of instabilities of gravitationally bound objects \cite{add}. In
\cite{sno}, $f(R)$ has the form $a R^m + b R^{-n}$ (where $a, b,m,n$ are real
numbers). The $R$ term with a positive power is dominant in the early universe giving
power-law inflation for $1< m < 2$. The $R$ term with a negative power dominates in the
late universe, when curvature is small and gives late-time acceleration. In other
attempts adopting this approach, $f(R)$ has been taken as $ln R$ \cite{sns}
and $Tr(1/R)$ \cite{de}. Some other modifications are suggested
in \cite{smc,scv}. Recently, a function of Gauss-Bonnet term has
been considered for $f(R)$ as gravitational dark energy \cite{snsd}. In
\cite{ka}, $f(R)$ is taken as Yukawa-like term of $R$ and in \cite{om}, $f(R)
\propto (a R^2 + b R^{\mu\nu}R_{\mu\nu} +
cR^{\mu\nu\rho\sigma}R_{\mu\nu\rho\sigma})^{-1}$, which leads to current
acceleration without resorting to dark energy.  Thus, in
the past few years, many models 
have been proposed to explain accelerated universe. A review on modified gravity as an
alternative to dark energy is available in \cite{sn06}. A more comprehensive review is provided in
\cite{ejc06}.

All of these models are phenomenological in the sense that
an idea of dark energy is introduced $a$ $priori$ either in terms of gravitational
field or non-gravitational field. In spite of these attempts, cosmic
acceleration is still a challenge. So, a model is required that can explain early
inflation followed by a decelerated phase and then late acceleration beginning in the
recent past. Here, an attempt is made in this direction explaining fall of DED
from a very high value to current low value. The gravitational origin of dark energy is
discussed using an approach different from the approach in \cite{sc, sm, add,  sno,sns,de,smc, scv,snsd, sn06}, where non-linear term of curvature is
considered as dark energy lagrangian. In recent work \cite{lds}, it has been shown that
$f(R)$- dark energy models do not give viable cosmology. The model considered here is not $f(R)$- dark
energy model, but it is a cosmological model based on $f(R)$-gravity. There is
a crucial difference between these two. In the former case, which is
criticized in \cite{lds}, non-linear terms of curvature are treated as dark
energy. By contrast in the latter case, neither linear nor
non-linear term is considered as dark energy.In the present model,  dark energy emerges
from linear term
(Einstein-Hilbert term) and non-linear term $\alpha R^{(2 - r)}$ ($\alpha$ is
a coupling constant) of the scalar curvature $R$ in the action.   DE, obtained
here, mimics quintessence. Although it is
 induced by the gravitational sector, a corresponding scalar $\phi$ is
 derived here. In the following section, it will be shown that $r = 0.75 (1 + {\rm w}_{\rm de})$ with ${\rm w}_{\rm de}$ being the
equation of state parameter for dark energy In the case of quintessence DE, $ -
1/3 > {\rm w}_{\rm de} > - 1,$ so $ 0 < r < 1/2$ yielding $2 > (2 - r) >
3/2$. As $(2 - r) > 0, $ the term $\alpha R^{(2 - r)}$ does not cause
instability problem mentioned previously \cite{sc, sm, add}.

 In the present model, the  Friedmann  equation is derived with dark energy ,
 dark radiation and  dark matter density 
 emerging from the gravitational sector  as 
the combined effect of curvature and a scalar field $\chi$. Spontaneous emergence
 of dark energy from the gravitational sector, which is a physical concept, is caused by  dual roles
 of $R$ as a physical as well as a geometrical field  \cite{sks05, ks, aas,
 skp} .   

This leads to a very interesting cosmology. Here, cosmological picture begins
 with power-law 
 inflation, driven by curvature-induced dark energy, for a fraction of a second and universe expands $10^{27}$
 times. It is shown that elementary particles are created  
 due to decay of the curvature-induced dark energy scalar $\phi$ during the
 inflationary period being dubbed phase I. Radiation is emitted due
 to decay of fluctuation of $\phi$ at the end of this phase. This radiation
 heats the universe and existing elementary
 particles attain thermal equllibrium with emitted radiation
 quickly. Temperature of this radiation is evaluated to be $1.4\times 10^{17}
 {\rm  GeV}.$ This event heralds the big-bang scenario giving a possible geometrical
 origin of cosmic microwave background radiation. Thus, we have two kinds of radiation (i) dark
 radiation and (ii) emitted radiation due to decay of fluctuations of
 $\phi$. The emitted also thermalizes dark radiation and radiation component
 (including elementary particles and dark radiation in thermal equllibrium with
 emitted radiation)
 in Friedmann equation attains the temperature $1.4\times 10^{17}
 {\rm  GeV}.$ During the inflationary period, dark energy density falls by 3 orders. As a
 result, radiation  dominates  and universe
 decelerates. 

It is mentioned above that dark matter emerges as a combined
 effect of curvature and another scalar $\chi$. The dark matter, so obtained,
 is non-baryonic.  Apart from it, it is proposed
 that baryonic matter is formed  from created elementary particles during phase I
 (inflation) through various proceesses of standard cosmology such as
 nucleosynthesis, baryosynthesis and  hydrogen recombination, which are not
 discussed here. 

 According to Wilkinson Microwave Anisotropy Probe (WMAP) \cite{abl},
 around $386$  
 kyr of age, matter begins to dominate over radiation and deceleration
 continued . At 
 red-shift $Z_{**} \simeq 0.4$, a transition takes place. Dark energy re-dominated and
 decelerating universe got speeded up. Dark energy density, obtained here, falls from
 $2.58\times 10^{68} {\rm GeV}^4$ to its current 
 value falling by 115 orders. Thus, this model explains early and late cosmic
 acceleration including deceleration for a long period between these two
 phases consistent with standard cosmology results. Thus, this model  gives a unified
 picture of cosmological evolution. Interestingly, it is found below that this
 is the situation of 
 our observable universe, which resides in a larger unobsrvable universe.

The paper is organized as follows. Section 2.1 contains  action of the
theory, basic field equations 
and manifestation of curvature scalar as a physical field. Friedmann equation,
giving cosmic dynamics is derived in section 2.2. Scalar field models for dark energy
have 
been very popular. Though dark energy has gravitational origin here,  it is still
interesting to probe scalar field $\phi$and its corresponding potential that give rise to it.
 This probe is done in section 2.3. In section
3, Friedmann equation is solved for when dark energy drives the cosmic dynamics and it is
found that universe inflates with power-law expansion for an extremely short
period. The creation of elementary particles and radiation, due to decay of $\phi$
is also discussed in this section. Moreover, this section contains discussion on other three phases of the
universe.In section 4, idea 
of larger universe is introduced which is inspired by fall of DED by 115
orders. Results of the paper are summarized in
the last section.

Natural units ($\hbar = c = 1$) are used here with GeV as the fundamental
unit, where $\hbar$ and $c$ have their usual meaning. In this unit, it is
found that $1 {\rm GeV}^{-1} = 6.58 \times 10^{-25} {\rm sec}$.

\bigskip

\centerline{\underline{\bf 2. Dark energy Caused by dual role}}
 \centerline{\underline{\bf of the Ricci scalar curvature}}

\noindent{\underline{\bf 2.1 Dual role of the curvature scalar}}

Here, investigations begin with the action  
$$ S = \int {d^4x} \sqrt{- g} \Big[ - \frac{R}{16 \pi G} + \alpha R^{(2 - r)} +
\Big\{\frac{1}{2} g^{\mu\nu} \partial_{\mu}\chi \partial_{\nu}\chi - V (\chi)\Big\}
\Big], \eqno(2.1)$$
where $R$ is the Ricci scalar curvature, $G = M_P^{-2} ( M_P = 10^{19}$ GeV is
the Planck mass), $\alpha$ is a coupling constant with dimension (mass)$^{2r}$
with $r$ being a real number and $\chi$ is a scalar field with potential
$V(\chi)$. It is shown below that $(1 - r) > 0.$ So, the instability problem does
not arise. Here, a scalar field $\chi$ with potential $V(\chi)$ is introduced,
which yields dark matter density.

The action (2.1) yields gravitational field equations
$$- \frac{1}{16 \pi G} \Big[ R_{\mu\nu} - \frac{1}{2} g_{\mu\nu} R \Big] +
\alpha  [( 2 - r)\{
\triangledown_{\mu} \triangledown_{\nu}R^{(1 -r)} - 2 g_{\mu\nu} {\Box} R^{(1
  -r)}  + $$ 

$$( 2 - r) R^{(1 -r)} R_{\mu\nu}\} - \frac{1}{2} g_{\mu\nu} R^{(2 -r)} ]
 + \frac{1}{ 2} \Big[\partial_{\mu}\chi \partial_{\nu}\chi - g_{\mu\nu}\Big(\frac{1}{2}  \partial^{\sigma}\chi \partial_{\sigma}\chi - V (\chi)\Big)\Big] = 0 \eqno(2.2)$$
using the condition $\delta S_g/{\delta g^{\mu\nu}} = 0.$ Here,
$\triangledown_{\mu}$
denotes covariant derivative and the operator $\Box$ is given as
$${\Box} = \frac{1}{\sqrt{-g}} \frac{\partial}{\partial x^{\mu}}
\Big(\sqrt{-g} g^{\mu\nu} \frac{\partial}{\partial x^{\nu}} \Big) $$
with $\mu, \nu = 0,1,2,3$ and $g_{\mu\nu}$ as metric tensor components.

Taking trace of eqs.(2.2), it is obtained that
\begin{eqnarray*}
{\Box}R &=& \frac{r}{R} \triangledown^{\mu}R \triangledown_{\mu}R +
\frac{1}{3(2 - r)(1 - r)} \Big[ \frac{R^{(1 + r)}}{16 \pi G \alpha} \\&& -
\frac{R^r}{2\alpha} \{- \partial^{\sigma}\chi \partial_{\sigma}\chi + 4 V
(\chi) \} - r R^2 \Big]
\end{eqnarray*}
\vspace{-1.8cm}
\begin{flushright}
(2.3)
\end{flushright}
with $ \alpha  > 0$ to avoid the ghost problem.

Eq.(2.3) is typically of the Klein - Gordon form with the source term
given on the right hand side of this equation. It shows that the Ricci scalar $R$ behaves as a physical field in addition to its usual role as a geometrical
field \cite{ks, aas, skp,sks04}.

Moreover, from the action(2.1), equation for $\chi$ is obtained as
$$ {\Box} \chi + V^{\prime} (\chi) = 0 \eqno(2.4a)$$
with $V^{\prime} (\chi) = dV/d\chi$ and components of energy-momentum tensor
as
$$ T^{\mu\nu}_{\chi} = \partial^{\mu}\chi \partial^{\nu}\chi - 
g^{\mu\nu} \Big(\frac{1}{2}  \partial^{\sigma}\chi \partial_{\sigma}\chi - V
(\chi)\Big) . \eqno(2.4b)$$

\smallskip

\noindent{\underline{\bf 2.2 Emergence of dark energy }

Experimental evidences support spatially homogeneous flat model of the
universe \cite{ad}. So, the line-element, giving geometry of the universe, is 
taken as
$$ dS^2 = dt^2 - a^2(t)[dx^2 + dy^2 + dz^2] \eqno(2.5)$$
with $a(t)$ as the scale factor.

In the space-time, given by eq.(2.5), eqs.(2.3) and (2.4a) are obtained as
$$ {\ddot R} + 3 \frac{\dot a}{a}{\dot R} - \frac{r {\dot R}^2}{R} =
\frac{1}{3(2 - r)(1 - r)} \Big[ \frac{R^{(1 + r)}}{16 \pi G \alpha} + $$
$$\frac{R^r}{2\alpha} \{- {\dot \chi}^2 + 4 V (\chi) \} - r R^2 \Big]  \eqno(2.6)$$
and
$${\ddot \chi} + 3 \frac{\dot a}{a} {\dot \chi} + V^{\prime}(\chi) = 0 \eqno(2.7)$$
as $\chi(x,y,z,t) = \chi(t)$ due to spatial homogeneity.  Here, overdot gives derivative with respect to time $t$.

In most situations ( matter-dominated model and dark
energy dominated models), we have $a(t)$  as a power-law solution, for which
$R$ is obtained as power-law function of $a(t)$. So, it is reasonable to take
$$ R = \frac{A}{a^n} ,  \eqno(2.8)$$
where $n \ne 0$ is a real number and $A$ is a constant with dimension
(mass)$^2$.

Using eq.(2.8), eq.(2.6) yields
$$\frac{d}{dt}\Big(\frac{\dot a}{a}\Big) + [3 - n (1 - r)] \Big(\frac{\dot
  a}{a}\Big)^2 = \frac{1}{3 n (2 - r)(1 - r)} \Big[- \frac{a^{-nr}}{16 \pi
  G \alpha A^r} - $$
$$\frac{1}{2\alpha } \Big(\frac{A}{a^n} \Big)^{(r - 1)} \{- {\dot \chi}^2 + 4
V (\chi) \} + r\frac{A}{a^n}  \Big]  \eqno(2.9)$$

Eq.(2.7) integrates to
$$\frac{d}{dt} \Big(\frac{1}{2} {\dot \chi}^2 + V(\chi) \Big) + 3 \frac{\dot
  a}{a} {\dot \chi}^2 = 0 . \eqno(2.10)$$

Moreover, in the space-time (2.5), energy density $\rho_{\chi}$ and ${\rm
  p}_{\chi}$ are obtained from eq.(2.4b) as
$$ \rho_{\chi} =  \frac{1}{2} {\dot \chi}^2 + V(\chi) \eqno(2.11a)$$
and
$${\rm  p}_{\chi} = \frac{1}{2} {\dot \chi}^2 - V(\chi). \eqno(2.11b)$$

For a cosmic fluid, pressure is proportional to its energy density. So, for a
cosmic fluid, given by the scalar field $\chi$,
$$ {\rm  p}_{\chi} = k \rho_{\chi} \eqno(2.11c)$$
with $k$ being the proportionality constant. From eqs.(2.11a), (2.11b) and
(2.11c), it is obtained that
$$ V(\chi) = \frac{[1 - k]}{ 2 [1 + k]} {\dot \chi}^2 .
\eqno(2.11d)$$

Now using $k = n (1 - r)/3$ in eq.(2.11d)
$$ V(\chi) = \frac{[3 - n (1 - r)]}{ 2 [3 + n (1 - r)]} {\dot \chi}^2 .
\eqno(2.12)$$

Connecting eqs.(2.12) and (2.10) and integrating, we obtain
$$ {\dot \chi}^2 = \frac{B}{a^{3 + n (1 - r)}} \eqno(2.13)$$
with $B$ being an integration constant with dimension (mass)$^4$.

Further, (2.12) and (2.13) yield
$$ - {\dot \chi}^2 + 4 V(\chi) = \frac{3 [1 - n (1 - r)]}{  [3 + n (1 -
  r)]}\frac{B}{a^{3 + n (1 - r)}}. \eqno(2.14)$$

Connecting eqs.(2.9), (2.11a,b), (2.12), (2.13) and (2.14), we have
$$\frac{d}{dt}\Big(\frac{\dot a}{a}\Big) + [3 - n (1 - r)] \Big(\frac{\dot
  a}{a}\Big)^2 = \frac{1}{3 n (2 - r)(1 - r)} \Big[- \frac{a^{-nr}}{16 \pi
  G \alpha A^r} - $$
$$ \frac{\{1 - n (1 - r) \}}{2 \alpha [3 + n (1 - r)]} \frac{B A^{(r - 1)}}{a^3} + r \frac{A}{a^n}  \Big]  \eqno(2.15)$$

In the space-time (2.5), the scalar curvature is calculated as
$$R = 6 \Big[ \frac{d}{dt}\Big(\frac{\dot a}{a}\Big) + 2 \Big(\frac{\dot
  a}{a}\Big)^2 \Big] .  \eqno(2.16a)$$

$R$, given by eqs.(2.8) and (2.16a) yield
$$6 \Big[ \frac{d}{dt}\Big(\frac{\dot a}{a}\Big) + 2 \Big(\frac{\dot
  a}{a}\Big)^2 \Big] = \frac{A}{a^n} .  \eqno(2.16b)$$

Elimination of $\frac{d}{dt}\Big(\frac{\dot a}{a}\Big)$ from eqs.(2.15) and
(2.16) leads to the Friedmann equation
$$ \Big(\frac{\dot a}{a}\Big)^2 =  \frac{1}{3 n (2 - r)(1 - r)[1 - n (1 - r)]} \Big[- \frac{a^{-nr}}{16 \pi
  G \alpha A^r} - $$
$$ \frac{\{1 - n (1 - r) \}}{2 \alpha [3 + n (1 - r)]} \frac{B A^{(r - 1)}}{a^3} \Big]  +
\frac{1}{[1 - n (1 - r)]} \Big(\frac{r}{3 n (2 - r)(1 - r)} - \frac{1}{6}
  \Big) \frac{A}{a^n}   \eqno(2.17)$$
giving dynamics of the universe.

Setting $n = 4,$ eq.(2.17) is re-written as
$$ \Big(\frac{\dot a}{a}\Big)^2 =  \frac{1}{12 (2 - r)(1 - r)[ - 3 + 4 r]} \Big[- \frac{a^{-nr}}{16 \pi
  G \alpha A^r} - $$
$$ \frac{\{ - 3 + 4 r) \}}{2 \alpha [7 - 4 r]} \frac{B A^{(r - 1)}}{a^3} \Big]  +
\frac{1}{6 [- 3 + 4 r]} \Big(\frac{r}{2 (2 - r)(1 - r)} - 1 \Big) \frac{A}{a^4}   \eqno(2.18)$$

It is interesting to see that, in eq.(2.18), the second term on the right hand
side ( being $\sim a^{-3}$ ) has the form of matter density $\rho_{\rm m}$ (
for pressureless baryonic and non-baryonic matter), which is the
combined effect of curvature and scalar field $\chi$ with potential given by
eq.(2.12). As this term emerges due to $\chi$ scalar in the action (2.1), it
is dark matter density $\rho_{\rm dm}$. The term $\sim a^{-4}$ has the form of
radiation density  emerging from curvature. Such a term appears
in brane-gravity induced Friedmann equation being  termed as dark
radiation \cite{rm}. So, in an analogous manner, here also,  the term proportional to
$\sim a^{-4}$ in (2.18) is recognized as dark radiation density $\rho_{\rm
  drd}$. The first term, on the right hand side of eq.(2.18) being
proportional to $a^{-4r}$ and  emerging from the gravitational sector, is
recognized as dark energy density $\rho_{\rm de}$.

It is reasonable to remark here that, in this approach, dark energy density
emerges due to presence of non-linear term of $R$ in the action (2.1). It is
unlike the approach in \cite{sc, sm, add, sno,sns, de,smc, scv,snsd, sn06},
where 
non-linear term of $R$ is used as lagrangian density for dark energy in the
action. Here, the non-linear term $R^{(2 - r)}$ is {\em not} dark energy
lagrangian.

Now, eq.(2.18) looks like
$$ \Big(\frac{\dot a}{a}\Big)^2 = \frac{8 \pi G}{3} (\rho_{\rm de} +
  \rho_{\rm  drd} +  \rho_{\rm dm} ), \eqno(2.19a)$$
where
$$ \frac{8 \pi G}{3} \rho_{\rm de} = - \frac{a^{-4r}}{192 (2 - r)(1 - r)[ - 3
  + 4 r] \pi   
G \alpha A^r} , \eqno(2.19b)$$
$$ \frac{8 \pi G}{3} \rho_{\rm  drd} = \frac{1}{6 [- 3 + 4 r]} \Big(\frac{r}{2
  (2 - r)(1 - r)} - 1 \Big) \frac{A}{a^4} \eqno(2.19c)$$
and
$$\frac{8 \pi G}{3} \rho_{\rm dm} = - \frac{(- 1 + 4 r)}{24 \alpha (2 -
  r)(1 - r)(7 - 4 r)} \frac{B A^{(r - 1)}}{a^3}. \eqno(2.19d)$$

Conservation equation for dark energy is given as
$$ {\dot \rho_{\rm de}} + 3 \frac{\dot a}{a} ( 1 + {\rm w}_{\rm de} )
\rho_{\rm de} = 0 , \eqno(2.20)$$
where  ${\rm w}_{\rm de} = p_{\rm
  de}/\rho_{\rm de}$ with $p_{\rm de}(\rho_{\rm de})$ being pressure(density)
for dark energy.

Connecting eqs.(2.19b) and (2.20), it is obtained that
$$ 4 r = 3 ( 1 + {\rm w}_{\rm de} ) . \eqno(2.21)$$

Using eqs.(2.21) in eqs.(2.19b,c,d), we obtain that
$$ \frac{8 \pi G}{3} \rho_{\rm de} = - \frac{a^{-3 ( 1 + {\rm w}_{\rm de} )
  }}{36 {\rm w}_{\rm de} (5 - 3{\rm w}_{\rm de} )(1 - 3{\rm w}_{\rm de}) \pi
  G \alpha A^r} , \eqno(2.22a)$$ 
$$ \frac{8 \pi G}{3} \rho_{\rm  drd} = \frac{1}{18 {\rm w}_{\rm de}}
\Big(\frac{3 ( 1 + {\rm w}_{\rm de} )}{4 (5 - 3{\rm w}_{\rm de} )(1 - 3{\rm
    w}_{\rm de})  } - 1 \Big) \frac{A}{a^4} \eqno(2.22b)$$ 
and
$$\frac{8 \pi G}{3} \rho_{\rm dm} = - \frac{8 (2 + 3 {\rm w}_{\rm de}) }{3
  \alpha  (5 - 3{\rm w}_{\rm de}
  )(4 - 3{\rm w}_{\rm de}) (1 - 3{\rm w}_{\rm de})} \frac{B A^{(2 + 3{\rm w}_{\rm de}
  )}}{a^3}. \eqno(2.22c)$$ 

Eqs.(22a,b,c) show that for positive $\rho_{\rm de}, \rho_{\rm  drd}$ and $\rho_{\rm dm}$

$${\rm w}_{\rm de} < - 2/3.$$

According to WMAP \cite{abl,am03}, current values of $\rho_{\rm dm}$ and $
\rho_{\rm de}$  are $\rho^0_{\rm dm} = 0.23 \rho^0_{\rm cr}$ and $ \rho^0_{\rm
  de} = 0.73 \rho^0_{\rm cr}$ with
$\rho^0_{\rm cr} = 3 H_0^2/{8\pi  G}, H_0 = 100{\rm h} {\rm km/Mpc sec} = 2.33
\times 0.68 \times 10^{-42} {\rm GeV}$ using ${\rm h} = 0.68$ ( a value having
the maximum likelihood ).

These observational values are used to determine constants in eqs.(2.22a) and
(2.22c). As a 
result,

$$ \rho_{\rm de}  = 0.73 \rho^0_{\rm cr} \Big(\frac{a_0}{a} \Big)^{3 (1 + {\rm w}_{\rm de} )} \eqno(2.23a)$$
and
$$ \rho_{\rm dm} = 0.23 \rho^0_{\rm cr} \Big(\frac{a_0}{a} \Big)^3, \eqno(2.23b)$$
where $a_0$ is the present scale factor.

Now, using eqs.(23a,b), Friedmann equation (2.19a) takes the form
$$\Big(\frac{\dot a}{a} \Big)^2 = H_0^2 \Big[
0.73  \Big(\frac{a_0}{a} \Big)^{3 (1 + {\rm w}_{\rm de} )} + 0.23
\Big(\frac{a_0}{a} \Big)^3 \Big] + \frac{8 \pi G}{3} \rho_{\rm  drd}. \eqno(2.24)$$

First two terms, on the right hand side of eq.(2.24) arising due to curvature
scalar $R$, are physical concepts, whereas term on the
left hand side is geometrical and it is caused by ${\Box} R$ in eq.(2.3). All these terms arise from the gravitational action
(2.1). It is due to manifestation of $R$ as a physical field as well as geometry\cite{sks05}.

\smallskip

\noindent{\underline{\bf 2.3 Scalar field for dark energy }

Here, dark energy is obtained from modified gravity. But, in non-gravitational models, it
is caused by some exotic matter given by a scalar field (quintessence, tachyon
and phantom). Though, we have a gravitational origin of DE here, it is natural to
probe a scalar $\phi$ (which is not a field for an exotic matter) giving
$\rho_{\rm de}$ obtained above. With $V(\phi)$ as potential, $\phi$ obeys the
equation
$${\ddot \phi} + 3 \frac{\dot a}{a}{\dot \phi} + V^{\prime}(\phi) = 0,
\eqno(2.25 )$$
and has mass dimension equal to 1. Here, $V^{\prime}(\phi) = d V/d\phi.$
DE density and pressure, in terms of $\phi$, are given as
$$ \rho_{\rm de} = \frac{1}{2} {\dot \phi}^2 + V(\phi) \eqno(2.26a)$$
and
$$ p_{\rm de} = \frac{1}{2} {\dot \phi}^2 - V(\phi) \eqno(2.26b)$$

It is interesting to see that conservation equation (2.20) yields (2.25) on
using (2.26a) and (2.26b).

Eqs.(2.26a) and (2.26b) yield
$$\rho_{\rm de} = \frac{{\dot \phi}^2}{1 + {\rm w}_{\rm de}} \eqno(2.27a)$$
and
$$ V(\phi) = \frac{1}{2} (1 - {\rm w}_{\rm de}) \rho_{\rm de} \eqno(2.27b)$$
using $p_{\rm de} = {\rm w}_{\rm de} \rho_{\rm de}$.
Now, from eqs.(2.21a) and (2.27a)
$${\dot \phi}^2 = \frac{2.19}{8 \pi G}(1 +{\rm w}_{\rm de}) H^2_0
\Big(\frac{a_0}{a} \Big)^{3 (1 +{\rm w}_{\rm de})}    \eqno(2.28a)$$
and
$$ V(\phi) = \frac{2.19}{8 \pi G}(1 - {\rm w}_{\rm de}) H^2_0
\Big(\frac{a_0}{a} \Big)^{3 (1 +{\rm w}_{\rm de})}    \eqno(2.28b)$$

\bigskip

\centerline{\underline{\bf 3. Cosmic evolution}}

\bigskip

In what follows, cosmic evolution passes through four different phases.

\smallskip

\noindent{\underline{\bf Phase 1. Power - law inflation}}

According to the present cosmological picture, in the beginning, the universe
contains DE only. So, Friedmann equation is given as
$$\frac{3 H^2}{8 \pi G} = \rho_{\rm de} = \frac{1}{2}{\dot \phi}^2 + V (\phi)
= \frac{{\dot \phi}^2}{1 + {\rm w}_{\rm de}}  \eqno(3.1)$$
using eqs.(2.26a) and (2.27a). Here $ H = {\dot a}/{a}$.

(2.24) looks like
$$\Big(\frac{\dot a}{a} \Big)^2 = 0.73 H_0^2 \Big(\frac{a_0}{a} \Big)^{3 (1 +
  {\rm w}_{\rm de} )} , \eqno(3.1)$$ 
 Eq.(3.3) yields
$${\dot \phi} = H \sqrt{\frac{3 (1 + {\rm w}_{\rm de} )}{8 \pi G}}
\eqno(3.2a)$$ 
and
$$\frac{3 H H^{\prime}}{4 \pi G} = \frac{1}{2} \frac{d {\dot \phi}^2}{d \phi}
+ V^{\prime}(\phi) = - 3 H {\dot \phi} \eqno(3.2b)$$
using eq.(2.25). Here, $H^{\prime} = dH/d\phi.$

Integrating eq.(3.2a), we obtain
$$ a = a(0) e^{\phi \sqrt{\frac{8 \pi G}{3 (1 +
  {\rm w}_{\rm de} )}}}  \eqno(3.3)$$
taking $\phi(0) = 0 $ without any harm to physics.

Connecting eqs.(2.28b) and (3.3)
\begin{eqnarray*}
V (\phi) &=& \frac{2.19}{16 \pi G} (1 - {\rm w}_{\rm de} ) H^2_0
  \Big[\frac{a_0}{a(0)} \Big]^{3 (1 +   {\rm w}_{\rm de} )} e^{-\phi \sqrt{\frac{4 \pi G}{5}}}
\end{eqnarray*}
\vspace{-1.8cm}
\begin{flushright}
(3.4)
\end{flushright}

Thus, we obtain dependence of the scale factor on $\phi$ during DE dominance
in the early universe. Using DE dominance, in the other form Friedmann
equation, given by (2.24), we obtain
$$\Big(\frac{\dot a}{a} \Big)^2 \simeq H_0^2
0.73  \Big(\frac{a_0}{a} \Big)^{3 (1 + {\rm w}_{\rm de} )}, \eqno(3.5)$$ 
which is integrated to
$$ a(t) = a(0) \Big[ 1 + \sqrt{0.73} H_0 \Big\{\frac{3 (1 + {\rm w}_{\rm de}
  )}{2} \Big\} \Big(\frac{a_0}{a(0)} \Big)^{3 (1 + {\rm w}_{\rm de} )/2} t \Big]^{2/3 (1 + {\rm w}_{\rm de} )} , \eqno(3.6)$$  
 where $a(0) = a(t=0) \ne 0,$ which is justified in what follows.

 At the
  beginning of the universe ,i.e. at $t = 0,$ energy density $\rho_{\rm de}$
  is expected to be very high, but {\em not} infinite. According to eq.(2.22a),
  $\rho_{\rm de}$ is {\em infinite}, if $a(0) = 0$. It suggests that $a(0)$
  should be non-zero to avoid the unphysical situation of divergent $\rho_{\rm
  de}$.

As ${\rm w}_{\rm de} < - 1/3, 3(1 + {\rm w}_{\rm de} ) > 2.$ Hence $a(t)$,
given by eq.(3.2), exhibits power-law inflation as ${\ddot a(t)} > 0.$

 Further, eqs.(3.2a) and (3.2b)
imply 
$$\frac{  H^{\prime}}{4 \pi G} = - H\sqrt{\frac{3 (1 + {\rm w}_{\rm de} )}{8
    \pi G}} \eqno(3.7)$$ 
giving the slow-roll parameter
$$\epsilon = \frac{1}{4 \pi G} \Big(\frac{H^{\prime}}{H} \Big)^2 = \frac{3 (1
  + {\rm w}_{\rm de} )}{2} \eqno(3.8)$$

Experimental results restrict $\epsilon \lesssim 0.05$ \cite{arl}. So, taking $\epsilon
\simeq 0.05, $ eq.(3.8) yields
$$ \frac{3 (1 + {\rm w}_{\rm de} )}{2} = \frac{1}{20}  \eqno(3.9a)$$
and
$$ {\rm w}_{\rm de} = - \frac{29}{30}. \eqno(3.9b)$$
It shows that DE, obtained here, mimics quintessence as $ {\rm w}_{\rm de} >
-1.$

 The maximum number of
e-fold for the inflation  \cite{arl}is
$$ N^{\rm max} = 63.3 + 0.25 ln \epsilon \simeq 62.55. \eqno(3.10)$$

If $a_c$ is the scale factor at the end of inflation
$$ \frac{a_c}{a(0)} = e^{62.55} = 1.46 \times 10^{27}. \eqno(3.11)$$

According to observational data, given above (in sub-section 2(b)), the
present DE density $\rho^0_{\rm de} = 2.18 \times 10^{-47} {\rm GeV}^4.$

In the present time,$\rho_{\rm de}$ acquired the value $\rho^0_{\rm de}$ (given above) falling by 115 orders from the beginning ,
$$ \rho(0) = 10^{115} \rho^0_{\rm de} =  2.18 \times 10^{68} {\rm GeV}^4.\eqno(3.12a)$$

(2.23a) yields
$$\rho_{\rm de}(0) = \rho_{\rm de}(t_*) \Big[\frac{a_*}{a(0)} \Big]^{3 (1 +
  {\rm w}_{\rm de} )} .\eqno(3.12b)$$
Using eqs.(3.9a), (3.11) and (3.12a), (3.12b) yields
$$\rho_{\rm de}(t_*) = 4.19 \times 10^{65} {\rm GeV}^4.\eqno(3.12c)$$
 
(2.27b) and (3.12c), it is obtained that
$$ V_c = 4.12 \times 10^{65} {\rm GeV}^4.\eqno(3.12d)$$

It is interesting to see that $V_c$ satisfies the observational constraint  \cite{arl}
$$ \frac{8 V_c}{3 M^4_P \epsilon} \simeq 2.6 \times 10^{-9}$$
for perturbation amplitude at the end of inflation.
where $ V_*$ is the potential at the end of inflation.
 From eqs.(2.23a) and (3.12a), it is obtained that
$$\Big[\frac{a_0}{a(0)} \Big]^{3 (1 +
  {\rm w}_{\rm de} )} = 10^{115} {\rm GeV}^4 \eqno(3.13a)$$
yielding an exceptionally large value of ${a_0}/{a(0)}$ as
$$ \frac{a_0}{a(0)} =  10^{1150} \eqno(3.13b)$$
For the time being, we postpone a discussion on this result to the next
section. 

Using the results (3.9a) and (3.13a), eq.(3.6) is obtained as
$$ a(t) = a(0) [ 1 + 1.52 \times 10^{14} t ]^{20}. \eqno(3.14)$$
Incorporation of the result (3.11) for $a_*$ in eq.(3.14), the inflation
period $t_c$ is evaluated as
$$ t_c \simeq 1.44 \times 10^{-13} {\rm GeV}^{-1} = 9.44 \times 10^{-38} {\rm
  sec} .  \eqno(3.15)$$

 At the end of inflation, it is obtained that
$$ \phi_c  = \sqrt{\frac{3 (1 +   {\rm w}_{\rm de} )}{8 \pi G}} ln
\Big( \frac{a_c}{a(0)} \Big) = 6.155 \times 10^{19} {\rm GeV}  \eqno(3.16)$$
connecting eqs.(3.3),(3.9b), (3.11) and (3.12d) . Here $\phi_c = \phi(t_c).$

\smallskip

\noindent {\underline{\bf 3.1.1 Particle creation during phase 1 and thermalization}}

\smallskip

It is obtained above that  $\phi$, given by (3.16), is the scalar for the
curvature induced quintessence dark energy. At the end of inflation, when
$\phi = \phi_c, V = V_c = 4.12 \times 10^{65} {\rm GeV}^4$ being given by (3.12d). Connecting (2.27b), (3.9b)
and (3.12a), it is obtained that at $t = 0, V = V_0$ is given as
$$ V_0 =  \frac{1}{2} (1 - {\rm w}_{\rm de}) \rho_{\rm de}(0) = 2.14 \times 10^{68} {\rm GeV}^4  \eqno(3.17)$$ 

These results show that $V(\phi)$ falls by $4$ orders, when $\phi$ rolls down
from the state $\phi = 0$ to $\phi = \phi_c$. During this period, early
universe inflates.

If $\Phi (t, {\bf x})$ and $\psi(t, {\bf x})$ are spinless field and spin-1/2
Dirac spinor present during this phase, it
is natural for these quantum fields to interact with curvature induced quintessence scalar
$\phi$ with interaction terms $ - \frac{1}{2} g \phi^2 \Phi^2$ and $- h {\bar
  \psi} \phi \psi ({\bar \psi} = \psi^{\dag} \gamma^0$ with
$\psi^{\dag}$ being adjoint of $\psi$ and $\gamma^0$ being time-component
of Dirac matrix in flat space-time. Here $g$ and $h$ are dimensionless
coupling constants.

\smallskip

\noindent \underline{\bf (i) Creation of spinless bosons}
\smallskip

 In the homogeneous
space-time (2,5), the equation (A1) (Appendix A) for $\phi$ with mass $m_{\Phi}$ and
interaction term with $\phi$ is obtained as
$$ {\ddot \Phi} + 3 \frac{\dot a}{a}{\dot \Phi} - a^{-2} \Big[
\frac{\partial^2}{\partial x^2} + \frac{\partial^2}{\partial y^2} +
\frac{\partial^2}{\partial z^2} \Big]\Phi (t, {\bf x})  + [g \phi^2(t)  +
m_{\Phi}^2] \Phi (t) = 0 .\eqno(3.18)$$

Using decomposition (A2) given in Appendix A, (3.18) is obtained as
$$ {\ddot \Phi} + 3 \frac{\dot a}{a}{\dot \Phi} + \Big[\frac{k^2}{a^2} + g
\phi^2(t)  + m_{\Phi}^2 \Big] \Phi (t, {\bf x}) = 0. \eqno(3.19)$$

Using the approximation 
$$ a(t) = = a(0) [ 1 + 1.52 \times 10^{14} t ]^{20}
\simeq  a(0) [ 1 + 3.04 \times 10^{15} t ]\eqno(3.20) $$ 
for the scale factor during phase I, conformal time $\eta$ is defined as
$$ \eta = \int^t {dt^{\prime}}/a(t^{\prime}) = [3.04 \times 10^{15} a(o)]^{-1}
ln [ 1 + 3.04 \times 10^{15} t ] \eqno(3.21)$$ 

Connecting (3.20) and (3.21),
$$ a({\tilde \eta}) \simeq a(0) e^{\tilde \eta}, \eqno(3.22a)$$ 
where
$${\tilde \eta} =  3.04 \times 10^{15} a(o) \eta . \eqno(3.22b)$$ 

Using the conformal time ${\tilde \eta}$ and defining $\Phi^c_k({\tilde \eta})
= a({\tilde \eta}) \Phi_k(t)$, (3.18) reduces to
$$ \frac{d^2 \Phi^c_k}{d {\tilde \eta}^2} + [{\tilde k}^2 - 1 + \{ {\tilde g}
\phi^2 + {\tilde m}^2_{\Phi} \} a^2(0) e^{2 {\tilde \eta}}]\Phi^c_k = 0,
\eqno(3.23)$$ 
where
$$ {\tilde k} = k/ 3.04 \times 10^{15} a(0), {\tilde k} = k/ [3.04 \times
10^{15} a(0)]^2 $$   and  
$$ {\tilde m}_{\Phi} = m_{\Phi}/3.04 \times 10^{15} a(0) .$$

Now two cases arise (i) when $\phi$ is very small i.e.  near to $ \phi
\simeq 0$ and (ii) when $\phi$ is large i.e. near to $ \phi
\simeq \phi_c$.

\noindent  \underline{\bf (a)Case I.$\phi$ is  near to $ \phi \simeq 0$ }

In this case, (3.23) reduces to
$$ \frac{d^2 \Phi^c_k}{d {\tilde \eta}^2} + [{\tilde k}^2  + {\tilde  m}^2_{\Phi} a^2(0) e^{2 {\tilde \eta}}]\Phi^c_k = 0, \eqno(3.24)$$ 
ignoring $1$ compared to other terms.

(3.24) yields the normalized solution as 
$$ \Phi^c_k = \Big[ \frac{\Gamma(1 + i {\tilde k}) \Gamma(1 - i {\tilde  k})}{2{\tilde k} (2\pi)^3} \Big]^{1/2} J_{ i {\tilde k}} \Big({\tilde
  m}_{\Phi} a(0) e^{\tilde \eta}\Big) . \eqno(3.25)$$ 
The normalization is done using the scalar product defined in (A6).

So, it is obtained that
\begin{eqnarray*}
 \Phi_k(t) &=& a^{-1}({\tilde \eta}) \Phi^c_k({\tilde \eta})\\& =& \Big[
    \frac{\Gamma(1 + i {\tilde k}) \Gamma(1 - i {\tilde  k})}{2{\tilde k}
    (2\pi)^3} \Big]^{1/2}  a^{-1}({\tilde \eta}) J_{ i  {\tilde k}} \Big({\tilde  m}_{\Phi} a(0) e^{\tilde \eta}\Big). 
\end{eqnarray*} 
\vspace{-1cm} 
\begin{flushright} 
(3.26)  
\end{flushright}

$in-$ and $out-$ states are obtained when ${\tilde \eta}$ tend to $- \infty$
and $\infty$ respectively. Thus 
$$ \Phi^{in}_k(t) = \Big[
    \frac{\Gamma(1 + i {\tilde k}) \Gamma(1 - i {\tilde 
    k})}{2{\tilde k} (2\pi)^3} \Big]^{1/2}  a^{-1}(-{\tilde \eta}) J_{ i
    {\tilde k}} \Big({\tilde  m}_{\Phi} a(0) e^{-{\tilde
    \eta}}\Big)  \eqno(3.27a)$$
and
$$ \Phi^{out}_k(t) = \Big[
    \frac{\Gamma(1 + i {\tilde k}) \Gamma(1 - i {\tilde 
    k})}{2{\tilde k} (2\pi)^3} \Big]^{1/2}  a^{-1}({\tilde \eta}) J_{ i
    {\tilde k}} \Big({\tilde  m}_{\Phi} a(0) e^{{\tilde
    \eta}}\Big). \eqno(3.27b)$$
 
Using definition of $\alpha_k$ and $\beta_k$ from (A8,9) and (3.27a,b), it is obtained that
$$ |\alpha_k|^2 = \frac{1 + {\tilde k}^2}{[ a^2(0) {\tilde k}]^{-2}}\quad{\rm and}\quad |\beta|^2 = [a(0) ]^{-4} .  \eqno(3.29a,b)$$ 
Condition (A10) and results (3.29a,b) yield 
$$ a^2(0) k = 1.  \eqno(3.30)$$ 

Non-zero value of $N_k = |\beta|^2$ shows creation of particle - antiparticle
spinless bosons \cite{lpark, ndb}.   

\noindent {\bf (b) Case 2. \underline{\bf $\phi$ is  near to $ \phi \simeq
  \phi_c$ }}
 
Proceeding in the same way, it is found that, in this case also, results for
$|\alpha_k|^2$ and $|\beta_k|^2$ are obtained as given in (3.29a,b).

The decay rate of $\phi$, due to creation of scalar particle-antiparticle
pairs \cite{lpark, ndb},  is given by
$$ \Gamma_{\phi \to {\bar \Phi}{\Phi}} = ln ({\prod_{K = -\infty}^{\infty}}|\alpha_k|^{-2})/V_4 = \frac{9.27
  \times 10^{15}a(0)}{(1 + a^4(0))}[ 1 + 1.52 \times 10^{14} t ]^{- 61} \eqno(3.31a)$$    
using (3.29a), (3.30) and 4-volume 
$$V_4 = \frac{a^3(0)}{9.27\times 10^{15}}[ 1 + 1.52 \times 10^{14} t ]^{61}.\eqno(3.31b)$$    

The decay can continue till decay rate $\Gamma_{\phi \to {\bar \Phi}{\Phi}}$
is larger than the expansion rate $H$ of the universe. So,
$$ \Gamma_{\phi \to {\bar \Phi}{\Phi}} \ge H.$$
It yields
$$\frac{9.27
  \times 10^{15}a(0)}{(1 + a^4(0))}[ 1 + 1.52 \times 10^{14} t ]^{- 61}  \ge
  \frac{3.04 \times 10^{15}}{[ 1 + 1.52 \times 10^{14} t ]}. \eqno(3.32a)$$
(3.32a) shows decay of $\phi$ to scalar particles upto the time
$$ t \simeq 6.7\times 10^{-15} \Big[\frac{a(0)}{1 + a^4(0)}\Big]^{1/60}. \eqno(3.32b)$$

\smallskip

\noindent \underline{\bf (ii) Creation of spin-1/2 Dirac fermions}
\smallskip

The action of Dirac field $\psi$ with interaction term $- h {\bar \psi} \phi
\psi$ and $m_{\psi}$ yields the equation
$$ i \gamma^{\mu} D_{\mu} \psi - [h \phi + m_{\psi}]\psi = 0  \eqno(3.33)$$
where $D_{\mu}$ is defined in (A11).

Using the conformal time $\eta$ (given by (3.21)), (3.33) is obtained as
$$[{\tilde \gamma}^0 \partial_0 + {\tilde \gamma}^1 \partial_1 + {\tilde
  \gamma}^2 \partial_2 + {\tilde \gamma}^3 \partial_3 + i (h \phi + m_{\psi}) a(\eta)]
  \Psi = 0,  \eqno(3.34)$$
where $\Psi = a^{3/2}(\eta)\psi.$

Here also, we take two cases as above.

\noindent {\bf (a) Case 1.} \underline{\bf $\phi$ is  near to $ \phi \simeq 0$ }

In this case, (3.34) reduces to  
$$[{\tilde \gamma}^0 \partial_0 + {\tilde \gamma}^1 \partial_1 + {\tilde
  \gamma}^2 \partial_2 + {\tilde \gamma}^3 \partial_3 + i m_{\psi} a(\eta)]
  \Psi = 0.  \eqno(3.35)$$

Operating (3.35) by  $[{\tilde \gamma}^0 \partial_0 + {\tilde \gamma}^1
  \partial_1   + {\tilde   \gamma}^2 \partial_2 + {\tilde \gamma}^3 \partial_3
  - i m_{\psi} a(\eta)]$ from left and using decompositions of $\Psi$ given in
  Appendix A, differential equation for $f_{k,s} (g_{k,s})$ is obtained as
$$ {\tilde f}^{\prime\prime} + [k^2 + i  m_{\psi} \epsilon a^{\prime}(\eta) +
m^2_{\psi}  a^2(\eta)]{\tilde f} = 0 , \eqno(3.36a)$$ 
where ${\tilde f} = f_{k,s} (g_{k,s}),{\tilde \gamma}^0 { u}_s =
  \epsilon { u}_s $, ${\tilde \gamma}^0 {\hat u}_s = \epsilon {\hat u}_s$ and
  $\epsilon = \pm 1$.

Introducing (3.22a,b), (3.36a) is re-written as 
$$ \frac{d^2 {\tilde f}}{d {\tilde \eta}^2} + [K^2 + i  {\tilde m}_{\psi} \epsilon a(0) e^{{\tilde \eta}}- {\tilde m}^2_{\psi} a^2(0) e^{2 {\tilde \eta}}
  ]{\tilde f} = 0 ,  \eqno(3.36b)$$
where $K = k/3.04\times 10^{15} $ and
  ${\tilde m}_{\psi} = - i m_{\psi}/3.04 \times 10^{15}.$

So, (3.36b) is approximated as
$$ \frac{d^2 {\tilde f}}{d {\tilde \eta}^2} + [K^2 - {\tilde m}^2_{\psi} a^2(0) e^{2 {\tilde \eta}}
  ]{\tilde f} = 0   \eqno(3.37)$$
ignoring $i  {\tilde m}_{\psi} \epsilon a(0) e^{{\tilde \eta}} =
  (m_{\psi}/3.04 \times 10^{15})\epsilon a(0) e^{{\tilde \eta}}$ compared to 
$(m_{\psi}/3.04 \times 10^{15})^2 \epsilon^2  a^2(0) e^{2{\tilde \eta}}$.

The normalized solution of (3.37) is obtained as
$$ f = \Big[\frac{\Gamma(1 + i K) \Gamma(1 - i K)}{ 2 K (2 \pi)^3}
  \Big]^{1/2} J_{iK} ( m_{\psi} a(0) e^{\tilde \eta}/3.04 \times
  10^{15}). \eqno(3.38) $$

Connecting (A14) and (A15) to (3.38), it is obtained that
\begin{eqnarray*}
\psi_{Ik,s} &=& a^{-3/2} \psi_{Ik,s} \\  &=& \Big[\frac{\Gamma(1 + i K) \Gamma(1 - i K)}{ 2 K (2 \pi)^3}
  \Big]^{1/2} a_0^{-3/2} e^{-3 {\tilde \eta}/2} \times \\&& J_{iK} ( m_{\psi}
  a(0) e^{ \tilde \eta}/3.04 \times 10^{15}) e^{- i {\vec
  k}.{\vec x}} { u}_s ,
\end{eqnarray*}
\vspace{-0.8cm}
\begin{flushright}
(3.39a)
\end{flushright}

\begin{eqnarray*}
\psi_{II(-k,-s)} &=& a^{-3/2} \psi_{II (-k,-s)} \\  &=& \Big[\frac{\Gamma(1 + i K) \Gamma(1 - i K)}{ 2 K (2 \pi)^3}
  \Big]^{1/2} a^{-3/2} \times \\&& J_{iK} ( m_{\psi} a(0) e^{ \tilde \eta}/3.04 \times 10^{15}) e^{i {\vec k}.{\vec x}} {\hat u}_{-s}.
\end{eqnarray*}
\vspace{-0.8cm}
\begin{flushright}
(3.39b)
\end{flushright}

Using $in$- and $out$-states defined as above we obtain $\psi^{in}_{Ik,s}$ and
$\psi^{out}_{I k, s}$ for ${\tilde \eta} < 0$ and ${\tilde \eta} > 0$
respectively. Similarly, $\psi^{in}_{II(-k,-s)}$ and
$\psi^{out}_{II(-k,-s)}$ are obtained.

So, definitions of $ \alpha_{k,s}$ and $ \beta_{k,s}$, given by (A 17) and (A
18) subject to the condition  (A 16), it is obtained that
$$ |\alpha_{k,s}|^2 = |\beta_{k,s}|^2 = \frac{1}{2}  \eqno(3.40a,b)$$
which shows creation of spin-1/2 particle-antiparticle pairs.

\noindent {\bf (b) Case 2.} \underline{\bf $\phi$ is  near to $ \phi \simeq
  \phi_c$ }

Like spinless case, here also,
$|\alpha_{k,s}|^2$ and $|\beta_{k,s}|^2$ are obtained as given in (3.40a,b)
for $\phi \simeq
  \phi_c$ .

The decay rate of $\psi$, due to creation of spin-1/2 particle-antiparticle
pairs \cite{lpark, ndb},  is given by 
   $$ \Gamma_{\phi \to {\bar \psi}{\psi}} = {\sum_{s = \pm 1}}ln ({\prod_{K =
   -\infty}^{\infty}}|\alpha_{k,s}|^{-2}/V_4 = \frac{1.854 \times 10^{15} ln
   2} {a^3(0)} [ 1 + 1.52 \times 10^{14} t ]^{61}   \eqno(3.41)$$
using $V_4$ given by (3.31b).

The decay of $\phi \to {\bar \psi}{\psi}$ can continue till decay rate $\Gamma_{\phi \to {\bar \psi}{\psi}}$
is larger than the expansion rate $H$ (obtained from (3.14)) of the universe. So,
$$ \Gamma_{\phi \to {\bar \Phi}{\Phi}} \ge H.$$
It yields that this decay will continue till 
$$ t \simeq 6.4 \times 10^{-15} {\rm GeV}^{-1}. \eqno(3.42)$$

\smallskip

\noindent \underline{\bf 3.1.2. Fluctuation of $\phi$ at the end of inflation  and thermalization} 
\smallskip

Here quintessence scalar is playing the role of inflaton causing power-law
inflation, so like inflaton, $\phi$ fluctuates as $ \phi = \phi_c(t) +
{\delta \phi(t,{\bf x})}$ at the end of inflation. 

From (2.25), Klein-Gordon equation for $\delta \phi(t,{\bf x})$ is obtained as
$${\ddot {\delta \phi(t,{\bf x})}} + 3 \frac{\dot a}{a}{\dot {\delta
    \phi(t,{\bf x})}} - a^{-2}\Big[\frac{\partial^2}{\partial x^2} + \frac{\partial^2}{\partial y^2} +\frac{\partial^2}{\partial z^2} \Big]{\delta \phi(t,{\bf x})}$$
$$+ V^{{\prime}{\prime}}(\phi_c) {\delta \phi(t,{\bf x})} = 0,
\eqno(3.43)$$

Using $ {\delta \phi(t,{\bf x})} = \phi_f(t) e^{-i {\bf k}. {\bf x}} $ in
(3.43 ), it is obtained that
$${\ddot  \phi_f} + 3 \frac{\dot a}{a}{\dot \phi_f} + \Big[\frac{k^2}{a^2} +
3.68 \times 10^{28} \Big] \phi_f = 0 \eqno(3.44a)$$

Using approximated $a(t)$ from (3.20) and $\tau = 1 + 3.04 \times 10^{15} t$,
(3.44a )is obtained as
$$ \frac{d^2 \phi_f}{d\tau^2} + \frac{3}{\tau} \frac{d\phi_f}{d\tau} +
\Big[\frac{k_c^2}{\tau^2} + 3.98\times 10^{-3} \Big] \phi_f = 0, \eqno(3.44b
)$$ where $k_c = k/3.04 \times 10^{15} a(0).$
This equation yields a solution
$$\phi_f = E \tau^{-1} J_{0.998} (0.063 \tau) .\eqno(3.44c)$$ 
Using the approximation
$$ J_n (x) \simeq \sqrt{2/\pi x} cos [x - \pi/4 + n \pi/2]$$
for large $x$ and definition of $\tau$ in (3.44c), it is obtained that
$$\phi_f = E (0.034 \pi)^{-1/2} [1 + 3.04 \times 10^{15} t]^{-3/2} cos [0.068
(1 + 3.04 \times 10^{15} t) - 1.499 \pi]. \eqno(3.45a)$$

So, 
$$  {\delta \phi(t,{\bf x})} = E (0.034 \pi)^{-1/2} [1 + 3.04 \times 10^{15}
t]^{-3/2}\times $$
$$ cos [ 2.07 \times 10^{14} t - 1.477 \pi] e^{-i {\bf k}. {\bf x}} \eqno(3.45b)$$
It shows that  $ {\delta \phi(t,{\bf x})}$ oscillates about $\phi = \phi_c$
with decaying amplitude. Decay of amplitude of a classical field shows loss of
energy from the system \cite{kof}. It means that we have two phenomena (i) production of
elementary particles and anti-articles of these during inflation and (ii) at
the end radiation of energy at the end of inflation. The radiated energy from
quintessence field $\phi$ heats up the universe and created particles set in
thermal equilibrium quickly if these particles have sufficient interaction.

\smallskip

\noindent {\underline{\bf Phase 2. Radiation-dominated phase}}

It is found above that , during phase 1, the universe inflates for a very
small fraction of a second and eqs.(3.12a) and (3.12c) show that $\rho_{\rm de}$
falls by 3 orders upto the end of inflation. The density for released dark energy is obtained as
$$ \rho_{\rm de (released)} = \rho_{\rm de} (0) - \rho_{\rm de}(t_c) =
2.175\times 10^{68} {\rm GeV}^4, \eqno(3.46)$$
using eqs.(3.12a) and (3.12c).

It means that due to decaying quintessence field $\phi$, DE emits as radiation
and elementary particles having thermal equllibrium with this radiation. It is parallel to the idea introduced by Bronstein
in 1933 \cite{brp}. Here, it is
proposed that this radiation causes cosmic microwave background radiation. So,
$$  \rho_{\rm de (released)}  = \frac{\pi^2}{15}
T_c^4. \eqno(3.47)$$

Initial temperature for this radiation is obtained as
$$ T_c = 1.35 \times 10^{17} {\rm GeV}  \eqno(3.48)$$
from eqs.(3.46) and (3.47).

We find another radiation term in Friedmann equation (2.19a) emerging
spontaneously from curvature along with DE term.

The emitted radiation, due to decay of $\phi$, sets dark radiation also in
thermal equilibrium. 

Moreover, from (3.46) and (3.47),it is obtained that
$$ \rho_{\rm rd} (t_c) > \rho_{\rm de}(t_c) . $$

Using the equation of state for emitted radiation $p_{\rm rd} = {\rho}_{\rm
  rd}/3$ (with $p_{\rm rd}$ and $\rho_{\rm rd}$ being the pressure and energy
density respectively)  conservation equation yields 
$$ \rho_{\rm rd}=  5\times 10^{-5} \rho^0_{\rm cr} \Big(\frac{a_0}{a} \Big)^4, \eqno(3.49)$$
using $\rho^0_{\rm rd} = 5\times 10^{-5} \rho^0_{\rm cr}$ where $\rho^0_{\rm cr} =
2.33 \times 0.68 \times 10^{-42} {\rm GeV}$ . Here $a_0$ is the present value
of the scale factor.

Emitted radiation and produced particles (due to decay of  quintessence DE)
also contribute to the energy density in FE (2.19a). As a result, (2.19a) is
modified as
$$ \Big(\frac{\dot a}{a}\Big)^2 = \frac{8 \pi G}{3} \Big[\rho_{\rm de} + 5\times 10^{-5}
  \rho^0_{\rm cr} \Big(\frac{a_0}{a} \Big)^4 +  \rho_{\rm drd}  +  \rho_{\rm
  dm} \Big], \eqno(3.50a)$$ 
where dark radiation density is given by
$$ \frac{8 \pi G}{3} \rho_{\rm  drd} = \frac{0.05A}{a^4} \eqno(3.50b)$$ 
being obtained from (2.22b) and (3.9b) for ${\rm w}_{\rm de}$.

$ \rho_{\rm dm}$ decreases as $a^{-3}$ with the scale factor. So, in the early
universe when $a(t)$ is very small radiation terms dominate over
matter. Moreover, (3.49) shows dominance of radiation over DE term when $t >
t_c$. Thus (3.50a) is approximated as
$$ \Big(\frac{\dot a}{a}\Big)^2 \simeq \frac{8 \pi G}{3} \Big[5\times 10^{-5}
  \rho^0_{\rm cr} \Big(\frac{a_0}{a} \Big)^4 + \frac{6.27\times 10^{35}
  A}{a^4}\Big]  .\eqno(3.50c)$$ 
As dark radiation also has thermal equilibrium with radiation emitted due to
  dark energy decay , (3.50c) reduces to 
$$ \Big(\frac{\dot a}{a}\Big)^2 \simeq \frac{8 \pi G}{3} \Big[5\times 10^{-5}
  \rho^0_{\rm cr} \Big(\frac{a_0}{a} \Big)^4 + \frac{6.27\times 10^{35}
  A}{a^4}\Big] = 5 \times 10^{-5} H_0^2 \Big(\frac{a_0}{a} \Big)^4 , \eqno(3.51)$$  
which integrates to
$$ a(t) = a_* [ 1 + 2 \beta (t - t_c) ]^{1/2}, \eqno(3.52)$$
with
$$ \beta = \sqrt{5\times 10^{-5}} H_0 \Big(\frac{a_0}{a_*} \Big)^2.$$

In terms of $\phi$ eq.(3.51) can be re-written as
$$ {\dot \phi}^2 \Big(\frac{1}{a} \frac{da}{d\phi}\Big)^2 = 5\times 10^{-5}
H_0^2 \Big(\frac{a_0}{a} \Big)^4 .  $$

So,
$$\Big(\frac{1}{a} \frac{da}{d\phi}\Big)^2 = \frac{5\times 10^{-5} H_0^2}{(1 +
  {\rm w}_{\rm de}) \rho_{\rm de}} \Big(\frac{a_0}{a} \Big)^4 = 1.77 \times
  10^{-42}\Big(\frac{a_0}{a} \Big)^{39/10}  \eqno(3.53)$$
using eq.(2.27a) and ${\rm w}_{\rm de} = - 29/30$. 

Eq.(3.53) is integrated to
$$ a(t) = a_c \Big[1 + 2.56 \times 10^{-20} \Big(\frac{a_0}{a_*} \Big)^{39/20}
(\phi - \phi_c) \Big]^{20/39}. \eqno(3.54)$$

Connecting eqs.(2.28b) and (3.54), it is obtained that
$$ V(\phi) = 4.3 \times 10^{-47}\Big(\frac{a_0}{a_*} \Big)^{1/10} \Big[1 + 2.56 \times 10^{-20} \Big(\frac{a_0}{a_*} \Big)^{39/20}
(\phi - \phi_c) \Big]^{- 2/39}. \eqno(3.55)$$

\smallskip

\noindent {\underline{\bf 3.3. Phase 3. Matter-dominated phase}}

Like radiation, we have two types of matter (i) dark matter given by
$\rho_{\rm dm}$ in (2.19a), which is non-baryonic and (ii) baryonic matter formed by elementary
particles, produced during phase I (inflation) through various processes of
standard cosmology such as nucleosynthesis, baryosynthesis and recombination
of hydrogen, which will be discussed in some other paper. Matter, so produced, contributes the density $\sim 0.04 \rho^0_{\rm cr}$ to
the present matter density $ \rho^0_{\rm m} \sim 0.27 \rho^0_{\rm cr}$  as $
\rho^0_{\rm dm} \sim 0.23 \rho^0_{\rm cr}$.

Matter dominates over radiation, when $\rho_m > \rho_r$, which implies that
$$ 0.27 > 5 \times 10^{-5} \Big(\frac{a_0}{a} \Big) \eqno(3.56)$$
using $\rho_m $ and $ \rho_r$ from eqs.(2.23b) and (3.49) respectively.

Defining red-shift as 
$$ 1 + z = \frac{a_0}{a}, \eqno(3.57)$$
the inequality (3.27) yields that matter dominates over radiation at 
$$ z < 5400 .$$

It is consistent with WMAP \cite{abl}, which gives that matter decouples from radiation at time $t_d = 386 {\rm
  kyr} = 1.85 \times 10^{37} {\rm GeV}^{-1}$ and red-shift 
$$ Z_d = \frac{a_0}{a_d} - 1 = 1089. \eqno(3.58)$$

So, we take the red-shift $ Z_d = 1089$, when matter dominates over
radiation. In this case, on the effective
Friedmann equation (2.24) is approximated as
$$\Big(\frac{\dot a}{a} \Big)^2 = 0.27 H_0^2 \Big(\frac{a_0}{a} \Big)^3 ,
\eqno(3.59)$$ 
including the contribution of created matter. (3.59) integrates to
$$ a(t) = a_d [ 1 + 2 \gamma (t - t_d) ]^{2/3}, \eqno(3.60)$$
with
$$ \gamma = \sqrt{0.27} H_0 \Big(\frac{a_0}{a_d} \Big)^{3/2} = 2.96 \times
10^{-38} {\rm GeV}$$
using eq.(3.58).

In terms of $\phi$, eq.(3.59) can be re-written as
$$ {\dot \phi}^2 \Big(\frac{1}{a} \frac{da}{d\phi}\Big)^2 = 0.27
H_0^2 \Big(\frac{a_0}{a} \Big)^3 .  $$

So,
$$\Big(\frac{1}{a} \frac{da}{d\phi}\Big)^2 = \frac{0.27 H_0^2}{(1 +
  {\rm w}_{\rm de}) \rho_{\rm de}} \Big(\frac{a_0}{a} \Big)^3 = 9.3 \times
  10^{-37}\Big(\frac{a_0}{a} \Big)^{29/10}  \eqno(3.61)$$
using eq.(2.27a) and ${\rm w}_{\rm de} = - 29/30$. 

Eq.(3.32) is integrated to
$$ a(t) = a_d [1 + 3.54 \times 10^{-14} (\phi - \phi_d) ]^{20/29}. \eqno(3.62)$$

Connecting eqs.(2.28b) and (3.33), it is obtained that
$$ V(\phi) = 4.33 \times 10^{-47} [1 + 5.5 \times 10^{-23}(\phi - \phi_d) ]^{- 2/29}. \eqno(3.63)$$

Further, formation of massive stars takes place during the matter-dominated
phase of the universe. Large structure formation is not addressed  in this work, but it is interesting to know the time , $t_B$, when the presence of massive stars is possible in this model. This time can be calculated from observations of Gamma-ray bursts (GRBs) taking place in sub-luminous star-forming host galaxies. GRBs show the existence of massive stars. According to previous observations, GRBs are found at high red-shift $z \sim 1.6$. But the recently observed GRB 050904 was found at red-shit $z_B = 6.1^{+0.36)}_{-0.12}$ beating all previous records. In what follows, $t_B$ is evaluated using this recent observation.

Connecting (3.57) and (3.60), $z_B$ is obtained as 
$$ 1 + z_B = \frac{a_0}{a(t_B)} =  \frac{a_0}{a_d} [1 + 4.44 \times 10^{-38}
(t_B - t_d)]^{-2/3}.  \eqno(3.64) $$

Using eq.(3.58) for $a_0/a_d$ and $z_B$ given above, (3.64) yields $t_B$ from 822 Myr to 918 Myr after big-bang showing that massive stars were present by this time.

\smallskip

\centerline{\underline{\bf 3.4 Phase 4. Dominance of dark energy in the late universe}}

Dominance of DE over matter ($\rho_{\rm de} > \rho_m$) is possible when
$$ 0.27 < 0.73 \Big(\frac{a_0}{a} \Big)^{-29/10} \eqno(3.65)$$
using $\rho_m $ and $ \rho_r$ from eqs.(2.23a) and (2.23b) respectively.

The inequality (3.65) shows that dominance of DE, in the late universe, is
possible, when
$$ \frac{a_0}{a} = 1 + z  < 1.409 \eqno(3.66)$$
using ${\rm w}_{\rm de} = -29/30$ obtained above.

So, we can take red-shift for the transition from $\rho_{\rm de} < \rho_m$ to $\rho_{\rm de} > \rho_m$  as 
$$ z_{**} = 0.4  \eqno(3.67)$$
safely. This is the red-shift at which DE dominates over
matter. As cosmic acceleration is caused by dominance of DE. $z_{**}$ is the
red-shift for transition from deceleration to acceleration.

Interestingly, it falls in the range of red-shift $z = 0.46 \pm 0.13$
for this transition, given by 16 Type supernova (SNe Ia) \cite{ag}. 

In this case, the
Friedmann equation (2.24) reduces to
$$\Big(\frac{\dot a}{a} \Big)^2 = 0.73 H_0^2 \Big(\frac{a_0}{a} \Big)^{1/10} , \eqno(3.68)$$ 
which integrates to
$$ a(t) = a_{**} [ 1 + \omega (t - t_{**}) ]^{20}, \eqno(3.69)$$
with
$$ \omega = 0.04 H_0 \Big(\frac{a_0}{a_{**}} \Big)^{1/20}\Big(\frac{a_0}{a_{**}}
  \Big)^{1/20} = 6.88 \times 10^{-44} {\rm GeV}$$
and $ a_{**} = a(t_{**}).$ Here $t_{**}$ is the transition time.
At the present time $t_0 = 13.7 {\rm Gyr} = 6.6 \times 10^{41} {\rm GeV}^{-1}$
we have
$$ 1 + z_{**} = \frac{a_0}{a_{**}} = [1 + 6.88 \times 10^{-44}(t_0 - t_{**})
]^{20} . $$

Connecting eqs.(3.67) and (3.69), it is evaluated that
$$ t_{**} = 4.13 \times 10^{41} {\rm GeV}^{-1} = 8.58 {\rm Gyr} . \eqno(3.70)$$

Eq.(3.69) shows accelerated expansion beginning at $ t_{**}$. So, $8.58 {\rm
  Gyr}$ is the time for transition from deceleration to acceleration in the
  late universe.

In terms of $\phi$, eq.(3.3) has the solution (for this case) as 
$$ a = a_{**} e^{[ 4 \sqrt{5 \pi G} (\phi - \phi_{**}) ]}  , \eqno(3.71)$$
where $\phi_{**} = \phi(t_{**}).$

Connecting eqs.(2.28b) and (3.71), it is obtained that
$$ V (\phi) = 2.22 \times 10^{37} e^{[ - \sqrt{4 \pi G/5} (\phi -
\phi_{**})]}. \eqno(3.72)$$

\bigskip

\centerline{\underline{\bf 4. Universe driven by dark energy}}

\bigskip

Eq.(3.12a) gives initial value of DE as $2.18 \times 10^{68}{\rm GeV}^4$. When
it falls to its current value $2.18 \times 10^{- 47}{\rm GeV}^4,$ the ratio of
$a_0$ (present scale factor) and $a(0)$ (the initial scale factor) is obtained
as $10^{1150}$ given by eq.(3.13b). It means that if universe is
derived by dark energy only, $ a_0 = 10^{1150} a(0)$.

 In this case, cosmic
dynamics is given by eq.(3.5) yielding the scale factor (3.6) and the  expansion rate as
$${\tilde H} = \frac{\sqrt{0.73} H_0 \Big(\frac{a_0}{a(0)} \Big)^{1/20}}{
  \Big[1 + \frac{1}{20} \sqrt{0.73} H_0 \Big(\frac{a_0}{a(0)} \Big)^{1/20} t
  \Big]}      \eqno(4.1)$$
with ${\rm w}_{\rm de} = - 29/30$ obtained above. 
Using eqs.(3.13a,b) in eq.(4.1)
$${\tilde H} \simeq \frac{20}{t} .   \eqno(4.2)$$

At $t=t_0 = 13.7 Gyr = 6.6 \times 10^{41} {\rm GeV}^{-1}$, eq.(4.2) yields
$$ {\tilde H_0} = 3.05 \times 10^{-42} {\rm GeV} = 19.25 H_0. \eqno(4.3)$$ 

It means that present Hubble's rate $H_0 \to {\tilde H_0} = 19.25 H_0$ in the
universe driven by DE only. It happens due to highly accelerating
expansion driven by DE having density (2.23a). The equation (4.3) shows that,
in this situation, $H$ is rescaled as $19.25 H = {\tilde H}$.  With this rescaling, eq.(3.1b) looks like
$$ \Big(\frac{\dot {\tilde a}}{\tilde a}\Big)^2 = \Big(19.25\frac{\dot  a}{
  a}\Big)^2 = 0.73 {\tilde H_0}^2 \Big(\frac{\tilde a_0}{\tilde a}\Big)^{1/10} \eqno(4.4)$$
and eq.(3.6) takes the form
$${\tilde a(t)} = {\tilde a(0)}\Big[1 + \frac{1}{20} \sqrt{0.73} H_0
  \Big(\frac{a_0}{a(0)} \Big)^{1/20} t \Big]^{20}, \eqno(4.5)$$
yielding 
$$\frac{\tilde a_0}{\tilde a(0)} \simeq \frac{a_0}{a(0)} = 5.23 \times
10^{1150}. \eqno(4.6)$$

This result is found for an universe, which is not being observed. We call it
{\it actual universe} Now we come
to our observable universe.

\smallskip
\underline{\bf 4.1.Observable universe}

As discussed in the preceding section,  universe passes through
4 different phases ( power-law inflation, radiation-dominated,
matter-dominated and DE dominated phase). We call it {\it observable
universe}. According to analysis for observable universe, eq.(3.11) gives the
ratio 
$$ \Big(\frac{a_c}{a(0)} \Big)_{\rm obs.} = 1.46 \times 10^{27} . \eqno(4.7)$$
At $t=t_c$ ($t_c$ being the time for end of inflation), eq.(3.49) yields
$$ \rho_r (t_c) = \frac{15 \times 10^{-5}}{8 \pi G} H_0^2  \Big(\frac{a_0}{a_c}
\Big)^4_{\rm obs.} . \eqno(4.8)$$

Connecting eqs.(3.46)-(3.49) and (4.8), we obtain that
$$\Big(\frac{a_0}{a_c} \Big)_{\rm obs.} = 6.43 \times 10^{29} . \eqno(4.9)$$

From eqs.(4.7) and (4.9)
$$\Big(\frac{a_0}{a (0)} \Big)_{\rm obs.} = 9.39 \times 10^{56}
. \eqno(4.10)$$

\smallskip
\underline{\bf 4.2. Actual universe and Observable universe}

So, from eqs.(4.6)and (4.10), it is obtained that
$$\Big(\frac{a_0}{a (0)} \Big)_{\rm actual} = 5.23 \times
10^{1150} >> \Big(\frac{a_0}{a (0)} \Big)_{\rm obs.} \eqno(4.11)$$

The result (4.11) prompts us to think that actually,
our observable universe resides in an extremely large universe driven by DE
only. As $H_0 \to {\tilde H_0} = 19.25 H_0$
$$ \rho_{\rm de(actual)} = 370.56 \rho_{\rm de(obs.)} . \eqno(4.12)$$
It means that only $\sim \frac{1}{371}$ of DE is available in our observable
universe. 

The reason for this scenario is given as follows. Due to the reverse gravity
effect of DE, the {\em actual universe} speeds up very fast and acceleration
continues upto the present epoch giving the ratio (4.6), whereas the {\em
  observable universe} (driven by DE) inflates for a very short period in the
beginning as well as decelerates driven by radiation and subsequently by matter
 for the major portion of its age. Moreover, in the recent past, a
transition from deceleration to acceleration takes place, which is consistent
with 16 Type SNIa observations \cite{ag} Thus, the bifurcation between the
{\em actual} and {\em observable} universe begins in early times, when
inflation of observable universe ends and its decelerated expansion, driven by
radiation as well as dark matter followed by acceleration in the recent past. As a result, for the observable universe, we have the ratio (4.10),
which is extremely less the similar ratio for the {\em actual universe} given
by (4.6). So, it is reasonable to conclude that the {\em observable universe}
is concentrated in a very small part of the {\em actual
  universe}. 

\bigskip

\centerline{\underline{\bf 5. Summary}}

\bigskip
It is interesting to see that, in the present model, DE emerges
spontaneously due to manifestation of the Ricci scalar curvature $R$ as a
physical field too. The physical aspect of $R$ is obtained adding a non-linear
term  to Einstein-Hilbert term.  In \cite{ sc,sm,add,sno,sns,de,smc,scv,snsd},
the non-linear 
term of $R$ stands for the lagrangian density of DE and it is added to explain
late acceleration in the universe. In \cite{sno}, Nojiri and Odintsov have
taken non-linear term $a R^m + b R^{-n}$ with the motive to have inflation in
the early universe and late acceleration, where $ R^m$ yields acceleration in
the early universe due to high curvature in the beginning of the universe and
$R^{-n}$ gives acceleration in the late universe due to low curvature there.
In contrast to this approach, here, only one non-linear term of $R$ is taken
in the action, which is not DE lagrangian. 
 Here, DE is induced by linear and non-linear terms of $R$ causing early and
 late acceleration. The resulting cosmological scenario is given as
 follows. In the beginning, our observable universe inflates for $\sim
 10^{-37} {\rm sec.}$ DE,  released during this period, decays to emission
of CMB radiation and light elementary particles. This event heralds usual
big-bang expansion with an end of inflation as emitted radiation dominates
over DE by the time $10^{-37} {\rm sec.}$ After this epoch, universe
decelerates for a long time, driven by radiation and subsequently by
matter. Around the age $8.58 {\rm Gyr}$ of the universe, a transition takes
place and DE re-dominates over matter, giving a cosmic jerk. As a result,
decelerating universe gets speeded up \cite{sk05}. DE density, obtained here,
falls from $2.58\times 10^{68}{\rm GeV}^4$ to its current value $2.18\times
10^{-47}{\rm GeV}^4$. Interestingly, it is found that the {\em observable
  universe}, having this scenario, expands within the {\em actual universe},
which is not observed.

In non-gravitational field theoretic models of DE, a scalar field $\phi(t)$
(quintessence, k-essence, tachyon and phantom ) are taken as DE source. In
contrast to this approach, here, $\phi(t)$ giving DE originates from gravitational sector. Thus, here, $\phi(t)$ has a geometrical origin,
whereas origin of quintessence, k-essence and phantom is not known (except
tachyon having stringy origin). The potential $V (\phi)$ ,for DE obtained
here, is tabulated below in different situations.

\bigskip

\centerline {\underline{\bf Table no.1}}  

\bigskip
{\sl Potential} $V (\phi)$ {\sl during different phases of the universe} 

\smallskip
$$
\begin{array}{lll}
\hline
V (\phi)  & {\rm Dominance}\quad {\rm in}   &  {\rm Nature}\quad {\rm  of} \\

         & {\rm the}\quad {\rm  universe}  & {\rm expansion}    \\
\hline
7.12 \times 10^{62} e^{-[\phi -
  \phi(0)]\sqrt{\frac{4 \pi G}{5}}}       &  {\rm   DE}   & {\rm acceleration} \\

 4.3 \times 10^{-47}\Big(\frac{a_0}{a_*} \Big)^{1/10} [1 +  &{\rm radiation} &
   {\rm deceleration} \\

2.56 \times 10^{-20}\Big(\frac{a_0}{a_*} \Big)^{39/20}
(\phi - \phi_*) ]^{- 2/39} & \omit & \omit\\

 4.33 \times 10^{-47} [1 +  &{\rm DM}
   & {\rm deceleration }\\
 5.5 \times 10^{-23}(\phi - \phi_d) ]^{- 2/29} & \omit & \omit\\

2.22 \times 10^{37} e^{[ - \sqrt{4 \pi G/5} (\phi -
\phi_{**})]} & {\rm DE}   &  {\rm acceleration} \\
\hline
\end{array}
$$

\smallskip

In \cite{jjh, ipn}, considering $V(\phi) \sim e^{-c \phi}$ ($c$ being a
constant) cosmic acceleration is derived in flat homogeneous model of the
universe. Here the result is opposite. In the present paper,
it is obtained that when universe accelerates , potential falls exponentially
and it falls as power-law in case of deceleration.

\bigskip

\centerline{\bf Appendix A}

\smallskip

\noindent \underline{\bf Boson case}

In the case of minimal coupling of $\Phi$ to gravity and having interaction
with $\phi$ as $-(1/2) g \phi^2 \Phi^* \Phi$, we have the equation

$$ {\Box} \Phi (t, {\bf x}) + [g \phi^2(t)  + m_{\Phi}^2] \Phi (t, {\bf x}) =
0 \eqno(A1)$$
with $m_{\Phi}$ being mass of $\Phi (t, {\bf x})$.

$\Phi$ is decomposed as
$$ \Phi (t, {\bf x}) = {\sum_{- \infty}^{\infty}} [\Phi_k (t) a_k
e^{i {\vec k}.{\vec x}} + \Phi^*_k (t) a_k^{\dag}
e^{-i {\vec k}.{\vec x}} ]. \eqno(A2)$$

In (A2), $a_k (a_k^{\dag})$ are  creation (annihilation) operators
satisfying quantum conditions
$$ [a_k , a_{k^{\prime}} ] = 0 = [a_k^{\dag} , a_{k^{\prime}}^{\dag} ]$$
$$ [a_k , a_{k^{\prime}}^{\dag} ] = (2 \pi)^3 \delta_{kk^{\prime}}
\eqno(A3,4,5) $$ 

The scalar product \cite{ndb} is defined as
$$ (\Phi_1, \Phi_2) = - i \int_{t = constant} \sqrt{-g} {d^3x}[ \Phi_1
{\partial}_t \Phi_2^* - \Phi_2^* {\partial}_t \Phi_1  ] \eqno(A6)$$
with $\partial_t$ denoting derivative with respect to time $t$ . 

The $in-$ and $out-$ states of $\Phi$ are obtained at two extremes of the
space-time where space-time is asymptotically Minkowskian. These are connected
through Boglubov transformations
$$ \Phi^{out}_k(t,x) = \alpha_k \Phi^{in}_k(t,x) + \beta_k \Phi^{in *}_k(t,x)
\eqno(A7)$$

This scalar product (A6) shows that $(\Phi^{in}_k, \Phi^{in *}_k) = 0$,
$(\Phi^{in }_k, \Phi^{in }_k) = 1 = (\Phi^{in *}_k, \Phi^{in
  *}_k)$. So 
From (A7), it is obtained that 
$$\alpha_k = (\Phi^{out}_k(t), \Phi^{in}_k(t)), \beta_k = (\Phi^{out}_k(t),
\Phi^{in {*}}_k(t)) \eqno(A8,9)$$
obeying the condition

$$|\alpha_k|^2 - |\beta|^2 = 1 .  \eqno(A10)$$

\smallskip

\smallskip

\noindent \underline{\bf Case of spin-1/2 Dirac fermion}

The equation of Dirac field $\Psi$ with mass $m_{\Psi}$ is given as
$$ [i \gamma^{\mu} D_{\mu} \Psi -  m_{\psi}]\Psi = 0  \eqno(A11)$$
with $i = \sqrt{-1}, D_{\mu} = \partial_{\mu} + \frac{1}{4} 
\Gamma^{\rho}_{\sigma\mu}h^{\sigma}_a)g_{\nu\rho}h^{\nu}_b {\tilde \gamma}^a
{\tilde \gamma}^.$ Here $\gamma^{\mu} = h^{\mu}_a{\tilde \gamma}^a (a =
0,1,2,3)$, $g^{\mu\nu} = h^{\mu}_a h^{\nu}_b \eta^{ab},$ 
$\{\gamma^{\mu}, \gamma^{\nu}\} = 2 g^{\mu\nu} $ and $\{{\tilde \gamma}^a,
{\tilde \gamma}^b = 2\eta^{ab}$ with $\gamma^{\mu} ({\tilde \gamma}^a)$
being Dirac matrices in curved(flat) space-time.

Further $\Psi$ are decomposed in discrete modes $k$ and spin $ (s/2)$ with $s =
\pm 1$ as
$$\Psi = \sum_{s = \pm 1} \sum_{k = - \infty}^{\infty} [b_{k,s} \Psi_{I k,s} +
d^{\dag}_{-k,s} { \psi}_{II k,s} ] \eqno(A12)$$
$$\Psi^{\dag} = \sum_{s = \pm 1} \sum_{k = - \infty}^{\infty} [
{\bar \Psi}_{I k,s} {\tilde \gamma}^0 b^{\dag}_{k,s} +
 {\bar \Psi}_{II k,s}{\tilde \gamma}^0 d_{-k,s}] \eqno(A13)$$
with $b_{k,s} (b^{\dag}_{k,s})$ and $d_{-k,s} (d^{\dag}_{-k,s})$ being
annihilation (creation) operators for positive (negative) energy particles
respectively. Further, using
$$\Psi_{I k,s} = f_{k,s}(\eta) e^{- i {\vec k}.{\vec x}} { u}_s,  \eqno(A14)$$
$$\Psi_{II k,s} = g_{k,s}(\eta) e^{i {\vec k}.{\vec x}} {\hat u}_s,  \eqno(A15)$$
where
$$
u_1 = \begin{pmatrix}
1 \\ 0 \\ 0 \\ 0 \end{pmatrix}, u_{-1} = \begin{pmatrix}
0 \\ 1 \\ 0 \\ 0 \end{pmatrix}, {\hat u}_1 = \begin{pmatrix}
0 \\ 0 \\ 1 \\ 0 \end{pmatrix}, {\hat u}_{-1} = \begin{pmatrix}
0 \\ 0 \\ 0 \\ 1 \end{pmatrix}. 
$$

Here also, $in$- and $out$- states of $\psi$ are obtained at two extremes of
the space-time where space-time is asymptotically Minkowskian.

The Bogolubov transformations for $\psi$  are given as
\begin{eqnarray*}
b^{out}_{k,s} &=& b^{in}_{k,s} \alpha_{k,s} + d^{in {\dag}}_{-k,-s}
\beta_{k,s} \\ (b^{out}_{-k,-s})^{\dag} &=&
\alpha^*_{k,s}(b^{in}_{k,s})^{\dag} + \beta^*_{k,s} d^{in}_{-k,-s} \\
(d^{out}_{-k,-s})^{\dag} &=& b^{in}_{k,s} \alpha_{k,s} + d^{in {\dag}}_{-k,-s}
\beta_{k,s} \\ d^{out}_{-k,-s} &=& \alpha^*_{k,s}(b^{in}_{k,s})^{\dag} + \beta^*_{k,s} d^{in}_{-k,-s}.
\end{eqnarray*}

The Bogolubov coefficients $\alpha_{k,s}$ and $\beta_{k,s}$ satisfy the
condition 
$$ |\alpha_{k,s}|^2 + |\beta_{k,s}|^2 = 1 ,  \eqno(A 16)$$
where
$$ \alpha_{k,s} = \int_{t = constant} d^3x \Psi^{in}_{I k,s} \Psi^{out
  {\dag}}_{I k,s} \eqno(A 17)$$
and
$$ \beta_{k,s} = \int_{t = constant} d^3x \Psi^{in}_{II -k,-s} \Psi^{out
  {\dag}}_{II -k,-s}. \eqno(A 18)$$

\end{document}